\documentclass[prd,twocolumn,floatfix]{revtex4}%
\pdfoutput=1 

\usepackage{amssymb,amsmath}
\usepackage{comment}

\def\simge{
    \mathrel{\rlap{\raise 0.511ex
        \hbox{$>$}}{\lower 0.511ex \hbox{$\sim$}}}}
\def\simle{
    \mathrel{\rlap{\raise 0.511ex
        \hbox{$<$}}{\lower 0.511ex \hbox{$\sim$}}}}
\def\beq{\begin{equation}}
\def\eeq{\end{equation}}
\def\beqa{\begin{eqnarray}}
\def\eeqa{\end{eqnarray}}
\def\bc{\begin{center}}
\def\ec{\end{center}}

\newcommand{\half}{$\frac{1}{2}$}
\newcommand{\threehalf}{$\frac{3}{2}$}
\newcommand{\fivehalf}{$\frac{5}{2}$}

\newcommand{\1}{ \bf 1}
\begin{document}


%
\title{Partial-wave and helicity operators for the scattering of two hadrons in lattice QCD}

\author{Stephen J. Wallace}
\email{stevewal@physics.umd.edu}
\affiliation{Department of Physics and Maryland Center for Fundamental Physics,
University of Maryland,
College Park, MD 20742, USA }

\begin{abstract}
Partial-wave operators for lattice QCD are developed in order to facilitate the identification of the spins of two-hadron scattering states corresponding to zero total momentum.  Taking the periodic boundary conditions for lattice states into account,  orthogonal sets of partial-wave operators for orbital angular momentum are identified.  When combined with the intrinsic spins of the hadrons,
orthogonal sets of parent operators for total angular momentum $J$ and projection $M$ are obtained.  The parent operators are subduced to irreducible representations of the octahedral group in order to obtain descendant operators for use in lattice calculations.  The descendant operators retain orthogonality with respect to $J$. The spin of a state can be identified by the spin of parent operators that dominate creation of the state.  For nonzero total momentum, operators are developed for a range of helicities and they are subduced to irreducible representations corresponding to the different directions of total momentum. Sets of operators that include a sufficient
 range of helicities allow identification of 
spin $J$ when a state couples to operators with helicities less than or equal to $J$, but not to operators with higher helicities.
\end{abstract}

\maketitle
\vspace{1cm} 
 

\section{Introduction  \label{sec:intro}}

Lattice operators subduced from irreducible representations (irreps) of group, 
$SU(2)$, to irreps of the octahedral group, $O_h$, have been used to determine the 
spins as well as the energies of single-hadron states in lattice QCD.  Recent works 
used such operators at zero momentum to determine the excited-state spectroscopy 
of mesons, based on two-quark operators~\cite{Dudek:2010wm}, and 
baryons, based on three-quark operators~\cite{Edwards:2011jj}.  

Subduction from $SU(2)$ associates each lattice operator with values of $J$ and $M$, 
where $J$ is the total angular momentum and $M$ is its projection along the z-axis. 

Using operators that are smeared as in Ref.~\cite{Peardon:2009gh}, rotational invariance is realized approximately for bound states that fit within the lattice volume, i.e.,
there is an approximate orthogonality 
between operators subduced from different values of $J$ and $M$.   
Using a different method to smear operators over many lattice sites, Ref.~\cite{Davoudi:2012ya}, finds
that the leading rotational-symmetry violating contributions from the 
finite lattice spacing, $a$, are 
suppressed by $\alpha_s a^2$ as $a \rightarrow 0$, where 
$\alpha_s = g_s^2/(4\pi)$ and $g_s$ is the 
strong coupling constant. 

The analyses of Refs.~\cite{Dudek:2010wm, Edwards:2011jj} obtain lattice correlation matrices that are 
approximately block-diagonal, with each block corresponding to 
a group of operators subduced from the same value of $J$.  Lattice excited states are found
to be created predominantly by operators subduced from a single value of $J$. That
identifies the spins of excited states created by single-hadron operators. 

However, multiparticle operators must be considered in view of 
the fact that 
 excited states of hadrons show up as resonances in
the scattering of two hadrons.  For example, the $\Delta$ baryon,
is a wide resonance in $\pi N$ scattering 
for spin $J = \frac{3}{2}$ and isospin $I = \frac{3}{2}$~\cite{Beringer:1900zz}. The quantum numbers match those of
a three-quark state in the quark model, or in lattice QCD.

In a well-known paper, L{\"u}scher\cite{Luscher:1990ux} has shown how 
phase shifts for the scattering of two hadrons can be extracted from lattice QCD.  
The L{\"u}scher 
method has been extended to systems with nonzero total momentum, unequal masses,
coupled channels and intrinsic spins~\cite{Rummukainen:1995vs, 
Bernard:2010fp, Hansen:2011mc, Leskovec:2012gb, Briceno:2012yi, 
Li:2012bi, Guo:2012hv}. 
It has been applied to coupled-channel scattering in 
Refs.~\cite{Dudek:2012gj,Dudek:2014qha,Wilson:2014cna}. 

In principle, the spins of scattering states can be identified by matching i), the pattern of approximately degenerate 
states found in lattice irreps with ii), the pattern of subductions of a given 
spin to the irreps of the lattice symmetry group. 
However, the combination of 
accidental degeneracies and statistical uncertainties can give 
ambiguous patterns of lattice states, e.g., 
the assignment of two or more values of $J$ may be compatible with them. 

In this work, sets of lattice operators 
based on partial waves are developed for 
the scattering of two hadrons with zero total momentum, and operators based on helicity are
developed for 
nonzero total momentum. The operators are designed to 
facilitate the identification of the spin in calculations of phase shifts. 

Periodic partial waves and their orthogonality properties are discussed in 
Section~\ref{sec:periodicPW}.  Operators based on partial waves are defined
in Section~\ref{sec:2Mops}.  
Using the partial-wave operators, intrinsic spins are included to obtain operators that create lattice 
states at zero total momentum corresponding to total spin $J$ in  Sec.~\ref{sec:P=0}.
These are called parent operators.  They are subduced to irreps of $O_h$ 
for use in lattice QCD.  The subduced operators are called descendant
operators.
Sets of mutually orthogonal parent operators are identified that correspond to a range of different total spins.

Section~\ref{subsec:P_not_0} develops parent operators based on helicity for nonzero 
total momenta. These are subduced to lattice irrep operators, the descendant operators, that
involve combinations of 
$\pm \lambda$.~\cite{Thomas:2011rh}
Sets of mutually orthogonal
operators are identified that correspond to a range of different helicities.  Section~\ref{sec:summary}
provides a brief summary of the work.
 
\section{Periodic partial waves \label{sec:periodicPW}}

Periodic boundary conditions in spatial directions apply to quark propagators on a lattice.   
Thus, lattice states created by quark-field operators generally are periodic in 
spatial directions. In spectroscopy based on single-hadron operators, the 
effects of the boundary conditions are reduced if the operators create 
states that fit approximately within the lattice volume.   
Scattering states are affected strongly by 
periodic boundary conditions.

Generally, scattering calculations incorporate the periodic boundary conditions
by using operators that involve combinations of plane-waves with  
 momenta restricted to ${\bf k} = \frac{2\pi {\bf n}}{L}$, where 
$L$ is dimension of the lattice in units of the lattice spacing and ${\bf n}$ is a vector with 
integer components along the axes of the lattice. 

Partial waves, $f_{\ell,m}({\bf x}) = j_{\ell}(k |{\bf x}|) Y_{\ell,m}({\hat x})$, 
involving momenta $ k = \frac{2\pi {|\bf n|}}{L}$, no longer obey the same 
periodicity as the plane waves. However, partial-wave operators create 
lattice states that must obey the periodic boundary conditions. 
Such lattice
states can be discussed in terms of periodic partial waves that are 
constructed from the formula,
\beq
\big[ f_{\ell,m}({\bf x})\big]^{(per)} = C \sum_{\bf n} f_{\ell,m}({\bf x}+{\bf n}L).
\label{eq:box_sum}
\eeq
The integer-valued components of ${\bf n}$ are limited by $-N \leq n_i \leq  N$ for i = 1 to 3
and the factor, $C = (2N+1)^{-3}$, provides a normalization.  
In the limit $N \rightarrow \infty$, the function $\big[f_{\ell m}({\bf x})\big]^{(per)}$ 
is periodic, 
\beq
\big[f_{\ell,m}({\bf x} + {\bf n}L)\big]^{(per)} = \big[ f_{\ell,m}({\bf x})\big]^{(per)},
\eeq
where ${\bf n}$ is any vector with integer components.  
  
Periodic partial waves for orbital angular momentum, $\ell$, and 
projection along the z-direction, $m$, have restricted orthogonality. 
Consider the overlap matrix, 
\beqa
\!\!\!\!\!\! O_{\ell^{\prime},m^{\prime};\ell, m}= 
\int d^3x &&~\bigr[j_{\ell^{\prime}}(k|{\bf x}|) 
       Y^{\dag}_{\ell^{\prime},m^{\prime}}({\hat x}\bigr]^{(per)}
\nonumber \\ 
&& \times \big[ j_{\ell}(k|{\bf x}|)Y_{\ell,m}({\hat x})\big]^{(per)} .
\label{eq:O-lm-orthog}
\eeqa
The integration symbol in Eq.~(\ref{eq:O-lm-orthog}) should be
read as a sum over $(L+1)^3$ lattice points that are inside or on the
boundaries of the $L^3$ lattice volume. 

In order to make diagonal overlaps equal to unity, define normalization factors
and construct the normalized matrix of overlaps as follows,
\beqa
&& N_{\ell, m} = \sqrt{ O_{\ell,m;\ell, m} }, \nonumber \\
&&  Z_{\ell^{\prime},m^{\prime};\ell, m}= \frac{1}{N_{\ell^{\prime},
m^{\prime}} N_{\ell, m}} O_{\ell^{\prime},m^{\prime};\ell, m}.
\label{eq:Z-lm-orthog}
\eeqa
Numerical calculations for finite lattice spacings show that,
\beqa
&& Z_{\ell^{\prime},m^{\prime};\ell, m} = 0 ,\nonumber \\
&& ~~if~~ \ell^{\prime} \neq \ell~(mod~2) ~and~  m^{\prime}~ \neq m~(mod~4),
 \label{eq:orthog-restrictions} 
\eeqa
holds to high precision.
The pattern of zero elements in Eq.~(\ref{eq:orthog-restrictions}) is the same as 
for L{\"u}scher's matrix, ${\cal M}_{\ell^{\prime},m^{\prime};\ell, m}$, that is central to the
method for determining phase shifts. As shown in Ref.~\cite{Luscher:1990ux},
the zeroes are exact because of cubic symmetry for a finite box with no discretization.
For a discrete lattice spacing, the orthogonality is found in
Appendix~\ref{app:l_m_orthog} not to depend
on the momentum in the spherical Bessel functions, or on
the difference between scattering wave functions and spherical Bessel functions.
In a lattice QCD calculation, the orthogonality should be realized 
approximately after averaging over gauge configurations.

\begin{table} 
\caption{ Pattern of even-parity matrix elements $Z_{\ell^{\prime},m^{\prime};\ell, m}$: 
${\bf 1}$ for diagonal elements, blank for zero elements and x for large off-diagonal
elements.  The top row shows the $\ell,m$ indices and the left column shows the $\ell^{\prime},m^{\prime}$
 indices. \label{tab:even-L-orthog}}
  \begin{tabular}{l|c|c|c|c|c|c|c|c|c|c|c|c|c|c|c|c|}
      \!&0,0\!&2,2\!&2,1\!&2,0\!&2,-1\!&2,-2\!&4,4&4,3&4,2\!&4,1\!&\!4,0\!&\!4,-1\!&\!4,-2\!&\!4,-3\!&\!4,-4\!\\
   \hline
  0,0& \1&   &   &   &   &    & x &   &   &   & x &   &   &   & x  \\ 
\hline
  2,2&   & \1&   &   &   & x  &   &   & x &   &   &   & x &   &    \\
\hline
  2,1&   &   & \1&   &   &    &   &   &   & x &   &   &   & x &    \\ 
\hline
  2,0&   &   &   & \1&   &    & x &   &   &   & x &   &   &   & x  \\
\hline
  2,-1&   &   &   &   &\1 &    &   & x &   &   &   & x  &   &   &       \\ 
\hline
  2,-2&   & x &   &   &   & \1 &   &   & x &   &   &   & x &   &     \\
\hline
  4,4& x &   &   & x &   &    & \1&   &   &   & x &   &   &   & x      \\ 
\hline
  4,3&   &   &   &   & x &    &   &\1 &   &   &   & x &   &   &     \\
\hline
  4,2&   & x &   &   &   & x  &   &   &\1 &   &   &   & x &   &     \\ 
\hline
  4,1&   &   & x &   &   &    &   &   &   &\1 &   &   &   & x &     \\
\hline
  4,0& x &   &   & x &   &    & x &   &   &   &\1 &   &   &   & x   \\ 
\hline
  4,-1&   &   &   &   & x &    &   & x &   &   &   &\1 &   &   &     \\
\hline
  4,-2&   & x &   &   &   & x  &   &   & x &   &   &   &\1 &   &     \\ 
\hline
  4,-3&   &   & x &   &   &    &   &   &   & x &   &   &   &\1 &     \\
\hline
  4,-4& x &   &   & x &   &    & x &   &   &   & x &   &   &   &\1   \\ 
\hline
\end{tabular}
\end{table}

There is one exception to the $mod$ 2 in Eq.~(\ref{eq:orthog-restrictions}): overlaps of periodic partial 
waves involving $\ell^{\prime},\ell$ = 0,2, or 2,0,
are zero. Thus, the $\ell$ = 0, 1 and 2 periodic partial waves are 
orthogonal to one another, except for the $m= \pm2$ values that are affected 
by the $mod$ 4 rule, i.e., the $\langle \ell^{\prime}, m^{\prime} | \ell, m \rangle$  
= $\langle 2,2|2,-2\rangle$ and $\langle 2,-2|2,2\rangle$ overlaps are of order 1. 

Periodic partial waves of opposite parities are orthogonal. 
Tables~\ref{tab:even-L-orthog} and \ref{tab:odd-L-orthog}
show the patterns of orthogonality of $Z_{\ell^{\prime},m^{\prime};\ell, m}$ for partial waves of the same parity,
using a blank to indicate a matrix element that is zero, ${\bf 1}$ to indicate
a diagonal element and $x$ to indicate a large off-diagonal element.

\begin{table}[h] 
\caption{ Pattern of odd-parity matrix elements $Z_{\ell^{\prime},m^{\prime};\ell, m}$. Symbols
are the same as in Table~\ref{tab:even-L-orthog}. The top row shows the $\ell,m$ indices and the 
left column shows the $\ell^{\prime},m^{\prime}$ indices.
\label{tab:odd-L-orthog}
}
  \begin{tabular}{l|c|c|c|c|c|c|c|c|c|c|c|c|c|c|c|c|}
     & 1,1& 1,0& 1,-1&3,3& 3,2& 3,1 & 3,0&3,-1&3,-2&3,-3 \\
   \hline
  1,1 & \1&   &   &   &   &  x &   &   &   & x \\
\hline
  1,0 &   & \1&   &   &   &    & x &   &   &   \\
\hline
  1,-1&   &   & \1& x &   &    &   & x &  &   \\
\hline
  3,3 &   &   & x & \1&   &    &   & x &   &   \\
\hline
  3,2 &   &   &   &   &\1 &    &   &   & x &   \\
\hline
  3,1 & x &   &   &   &   & \1 &   &   &   & x \\
\hline
  3,0 &   & x &   &   &   &    & \1&   &   &    \\
\hline
  3,-1&   &   & x & x &   &    &   &\1 &   &   \\
\hline
  3,-2&   &   &   &   & x  &   &   &   &\1 &   \\
\hline
  3,-3& x &   &   &   &   &  x &   &   &   &\1 \\
\hline
\end{tabular}
\end{table}

Periodic operators with $\ell$ = 0, 1 and 2 unavoidably couple to states of higher $\ell$ 
because of the cubic symmetry.  For example, a periodic operator for angular momentum $\ell$ = 1 couples
 with $\ell$ = 3, 5, $\cdots$ states.
Thus, assuming that $\ell \ge 5$ can be neglected, there is a need to distinguish which states are $\ell$ = 1 versus $\ell$ = 3.
The blank elements in Tables ~\ref{tab:even-L-orthog} and \ref{tab:odd-L-orthog}
show that it is possible to construct periodic 
operators for $\ell$ = 3 that are orthogonal to those for $\ell$ = 1. 
For example, $\ell,m$ = 3,2 is orthogonal to $\ell,m$ = 1,1, 1,0 and 1,-1. 
Thus, including a tracer operator that couples to states with $\ell \ge$ 3 
but does not couple to $\ell$ = 1 states, 
provides a means to distinguish lattice 
states involving $\ell = 3$ from those 
involving $\ell = 1$.  Similarly, states involving $\ell$ = 4 can be distinguished 
from ones involving $\ell$ =  2 by use of suitable tracer operators. 
That provides the motivation for the use of partial-wave operators corresponding
to distinct values of $\ell$ and $m$.  

\section{Plane-wave, lattice irrep and partial-wave operators \label{sec:2Mops}}

\subsection{ Plane-wave operators \label{subsec:PWops}}
 Consider two point-like fields for distinct, scalar mesons that are labeled by indices 1 and 2. 

The field operator for a single scalar meson at t=0, keeping only the creation part, is
\beq    
m^{\dag}_1({\bf x}, 0) =  \frac{1}{(2\pi)^{3/2}} \int d^{3}p \frac{1}{(2E_p)^{1/2}} a^{\dag}_{\bf p} e^{-i  {\bf p} \cdot  {\bf x}} . \eeq
Fourier transformation of the field gives the creation operator for a plane-wave state, 
\beqa   m^{\dag}_1({\bf k}, 0) &&= \frac{1}{(2 \pi )^{3/2}} \int d^{3}x ~e^{i {\bf k} \cdot {\bf x}}  m^{\dag}_1({\bf x}, 0), \nonumber \\
                       &&=  a^{\dag}_{\bf k} \frac{1}{(2 E_{\bf k})^{1/2}}. \eeqa
For total momentum and time both equal to 0, a two-meson operator is,
\beq     m^{\dag}_1({\bf k},0) m^{\dag}_2(-{\bf k},0) = a^{\dag}_{\bf k} a^{\dag} _{-{\bf k}}   \frac{1}{2 E_{\bf k}}. \eeq
  Acting on the vacuum, this operator creates a two-meson state with equal and opposite momenta,
\beq     m^{\dag}_1({\bf k},0) m^{\dag}_2(-{\bf k},0) |0\rangle  = |{\bf k},-{\bf k}\rangle \frac{1}{2 E_{\bf k}}.  \eeq
In a cubical volume of dimension $L$, the state is periodic when the momentum is ${\bf k} = \frac{2\pi {\bf n}}{L}$, where ${\bf n}$ is a vector 
whose components are integers. 

In the sink operator for two scalar mesons, destruction operators, $a_{\bf k'}$,  
are kept and in general the momentum of the sink operator, ${\bf k'}$,
has a different direction but the same magnitude as ${\bf k}$,
\beq   m_1({\bf k}',0) m_2(-{\bf k}',0) = a_{\bf k '} a_{-{\bf k}'} \frac{1}{2 E_{\bf k}}.  \eeq

Correlation functions involve the operators at $t=0$ and the Euclidean-time propagator, 
which is written as 
a sum over a complete set of eigenstates, $|n\rangle$, of the lattice Hamiltonian, as follows,
\beqa  
 C(t) = \sum_n \langle 0| m_1({\bf k}',0) m_2(-{\bf k}',0) |n\rangle   e^{-E_n t} 
\nonumber \\
\times \langle n| m^{\dag}_1({\bf k},0) m^{\dag}_2(-{\bf k},0) |0\rangle  ,
\nonumber 
\eeqa
yielding a correlation function for plane-wave states of two mesons, 
 \beq C(t) = \sum_n ~\langle {\bf k}', -{\bf k}'|n\rangle  e^{-E_n t} \langle n| {\bf k},-{\bf k}\rangle ~\frac{1}{4 E_{\bf k}^2} .
\eeq

\subsection{Lattice irrep operators }

  Lattice irrep operators involve sums of plane-wave operators over sets of 
momenta that are related to one another by rotations that leave the lattice invariant.
In general, the use of lattice irrep operators leads to mixings of various partial waves. 

An illustrative case is row 2 of the $T_1$ lattice irrep, for which the 
source operator is a sum of two terms,
 \beq   \frac{1}{2} m^{\dag}_1({\bf k},0) m^{\dag}_2(-{\bf k},0)   -  \frac{1}{2} m^{\dag}_1(-{\bf k},0) m^{\dag}_2({\bf k},0), \eeq
where ${\bf k}$ is along the z direction. The sink operator has a similar form 
except with annihilation operators.
The correlator can be written as, 
\beqa   C_{T_1,2} (t) &&= \frac{1}{4 E_{\bf k}^2} \sum_n \frac{1}{2}\Big(   
 \langle {\bf k}',-{\bf k}'| - \langle -{\bf k}',{\bf k}'| \Big) |n\rangle \nonumber \\
 && \times  e^{-E_n t}  \langle n| \frac{1}{2}\Big( |{\bf k},-{\bf k}\rangle  -
 |-{\bf k},{\bf k}\rangle \Big).
\eeqa
Inserting complete sets of states, $|{\bf x}_1, {\bf x}_2\rangle$, 
and using ${\bf x} = {\bf x}_1-{\bf x}_2$, ${\bf X} = \frac{1}{2}\big({\bf x}_1+{\bf x}_2\big)$
( similarly for ${\bf x}'$ and ${\bf X}'$), leads to 
 \beqa   C_{T_1,2} (t) \!\!\!\!\!\! &&= \frac{1}{4 E_{\bf k}^2} \sum_n e^{-E_nt} \nonumber \\
&& \times \int d^3 x_1' d^3 x_2' 
\frac{1}{2}\Big( e^{i {\bf k}' \cdot {\bf x}'} -  e^{-i {\bf k}' \cdot {\bf x}'} \Big)^{\dag}
\Psi_n ({\bf x}', {\bf X}') \nonumber \\
&& \times \int d^3 x_1 d^3 x_2 \Psi_n^{\dag} ({\bf x}, {\bf X}) 
 \frac{1}{2}\Big( e^{i {\bf k} \cdot {\bf x}} -  e^{-i {\bf k} \cdot {\bf x}} \Big),
\eeqa
where $ \Psi_n({\bf x},{\bf X}) = \langle {\bf x}_1,{\bf x}_2 |n\rangle$ is the
wave function of state $|n\rangle$.
For this example, partial-wave expansion of the factors involving plane-waves contain odd $\ell$ values, 
\beq 
\frac{1}{2}\Big( e^{i {\bf k} \cdot {\bf x}} -  e^{-i {\bf k} \cdot {\bf x}}\Big) = \sum_{odd~ \ell} 
4 \pi i^{\ell} j_{\ell}(k|{\bf x}|) Y^{\dag}_{\ell,0}({\hat k}) Y_{\ell,0}({\hat x}) .
\eeq
Thus, the correlator becomes, 
\beqa
C_{T_1,2}(t)\!\!\!\!\! && = \sum_n  e^{-E_nt}\frac{1}{4E_k^2}
\Big( \sum_{odd ~ \ell'} 4 \pi i^{\ell^{\prime}}
Y_{\ell',0}({\hat k}') \Phi_{n,\ell',0}(k)\Big)  
\nonumber \\
&& \times \Big( \sum_{ odd ~ \ell} 4 \pi i^{\ell} \Phi^{\dag}_{n, \ell,0}(k) Y^{\dag}_{\ell,0}({\hat k}) \Big),
\label{eq:Coft-1}
\eeqa
where 
\beq
 \Phi_{n,\ell,m}(k) = \int d^3x~  j_{\ell}(k|{\bf x}|) Y^{\dag}_{\ell,m}({\hat x}) \Phi_n({\bf x}),
\label{eq:Coft-2} \eeq
and
\beq
\Phi_n({\bf x}) = \int d^3 X \Psi_n({\bf x},{\bf X}) .
\label{eq:Coft-3}
\eeq
The matrix elements, $\Phi_{n,\ell,m}(k)$, show that the coupling of
 operators to lattice states involves the
$j_{\ell}(k|{\bf x}|) Y^{\dag}_{\ell,m}({\hat x})$ part of the operator.

When the mesons do not interact, the wave function, $\Phi_n({\bf x})$,  
corresponds to two mesons with the same momenta as in the operator that creates the state.
The wave function can  be expanded in partial waves as follows,
\beq
 \Phi_n({\bf x}) = \sum_{\cal{L,M}} a_{\cal{L,M}}~ j_{\cal L}(k|{\bf x}|)~Y_{\cal{L,M} }({\hat x}).
\label{eq:PWA}
\eeq
Substituting this expansion into Eq.~(\ref{eq:Coft-2}) and taking into account the
periodic boundary conditions leads to sums over the the overlaps of Eq.~(\ref{eq:O-lm-orthog}).

 When the two mesons interact, the scattering wave function generally differs from $j_{\cal L}(k|{\bf x}|)$.
Appendix~\ref{app:l_m_orthog} describes numerical tests that show that the interactions do not 
have a significant effect on the orthogonality.

\subsection{Lattice partial-wave operators}

In the analysis above, the two-meson operator at $t=0$ contains the plane-wave factor, 
$e^{i {\bf k}\cdot{\bf x}}$, and can be written as,
\beqa
&&\!\!\!\! m^{\dag}({\bf k},0)m^{\dag}(-{\bf k},0) = 
\nonumber \\
&& \!\!\!\! \frac{1}{(2\pi)^3}\int d^3 x_1 \int d^3 x_2 ~e^{i{\bf k}\cdot{\bf x}}~ m^{\dag}({\bf x}_1,0)m^{\dag}({\bf x}_2,0) .
\label{eq:mm_PW}
 \eeqa
This is a form that is useful for lattice QCD.
In order to restrict the source or sink operator to a single partial wave, 
the plane-wave factor is replaced by a partial-wave factor, defined as follows,
\beq
\Big[e^{i {\bf k} \cdot {\bf x}}\Big]_{\ell,m} \equiv  j_{\ell}(k|{\bf x}|)
Y_{\ell, m}({\hat x}) ,
\label{eq:PW} \eeq
where a factor $4 \pi i^{\ell} Y_{\ell,m}({\hat k})$ is omitted. The 
partial-wave operator is based on the factors that occur in the
matrix elements between operators and lattice states, as in
Eq.~(\ref{eq:Coft-2}).  The result is equivalent to projecting 
the plane-wave factor $e^{{\bf k}\cdot {\bf x}}$ in Eq.~(\ref{eq:mm_PW})
 to a partial wave using $Y_{\ell,m}(\hat{k})$.

 The partial-wave operator is defined by,
\beqa
&&\Big[m^{\dag}({\bf k},0)m^{\dag}(-{\bf k},0) \Big]_{\ell,m} = 
\nonumber \\
&& \frac{1}{(2\pi)^3}\int d^3 x_1 \int d^3 x_2 \Big[e^{i {\bf k} \cdot {\bf x}}\Big]_{\ell,m}
 m^{\dag}({\bf x}_1) m^{\dag}({\bf x}_2).~~
\label{eq:op_PW}
\eeqa
When used in a lattice QCD calculation, a partial-wave operator 
creates periodic states.

The quantization of momentum for noninteracting plane-waves  
that obey periodic boundary conditions implies that $q = \frac{kL}{2\pi} = |{\bf n}|$,
where ${\bf n}$ is a vector with integer components. Interactions modify
the wave functions but they should remain close to the noninteracting waves with momenta
corresponding to ${\bf n}^2$ = 1, 2, 3, 4 $\cdots$.  While 
there is no restriction that limits the momentum values, the
noninteracting momenta provide a reasonable choice for the momenta
in the partial-wave operators.  
The $j_{\ell}(k|{\bf x}|)$ factor can also be viewed as simply providing 
a spatial smearing with a smearing parameter, $k$, that may be chosen to
maximize the coupling of the operator to states of interest. 

A matrix of correlation functions could be obtained by using partial-wave operators 
with orbital angular momentum indices $\ell^{\prime}$ and $m^{\prime}$ 
at the sink point together with indices 
$\ell$ and $m$ at the source point, which gives,
\beqa
 C_{\ell^{\prime},m^{\prime};\ell,m}(t) = 
   \sum_n  e^{-E_nt} \Phi_{n,\ell^{\prime},m^{\prime}}(k) 
\Phi^{\dag}_{n, \ell,m}(k),
\eeqa
where $\Phi_{n, \ell,m}(k)$ is defined by Eq.~(\ref{eq:Coft-2}).
In this case, one can select operators corresponding to an 
orthogonal set of states, including tracer operators to indicate the presence of higher spins, 
in order to facilitate the identification of spins. However, the partial-wave 
operators are not used directly.  They serve as parent operators that 
are subduced to lattice irreps before use in lattice calculations.

\section{Operators for two-hadrons with intrinsic spins \label{sec:P=0}}

In lattice QCD, the representations of the intrinsic spins of hadrons are
orthogonal, like the spin states in quantum mechanics,
\beq
    \langle s, m_s | s^{\prime}, m_s^{\prime} \rangle = \delta_{s,s^{\prime}}\delta_{m_s,m_s^{\prime}} ,
\label{eq:s-ms_orthog}
\eeq 
where $s$ is the spin and $m_s$ is its projection along the z-azis.
Parent operators for the scattering of two hadrons with zero total momentum 
are constructed in Sec.~\ref{subsec:mB(0)}
by coupling the intrinsic spin angular momenta with a partial wave of orbital angular momentum 
to make total spin $J$ and projection $M$. 
  The final step, outlined in Sec.~\ref{subsec:subduce_P=0} is to produce
descendant operators by subducing the parent operators 
from $J$ and $M$ to lattice irreps of $O_h$. 

   When total momentum is nonzero, operators are constructed based on 
total helicity in Sec.~\ref{subsec:P_not_0}.  Lattice irrep operators are 
constructed in Sec.~\ref{subsec:P_not_0_irreps}.
 
\subsection{Parent operators for total momentum ${\bf P} = 0$ \label{subsec:mB(0)}}

Consider zero total momentum: ${\bf P} = {\bf k}_1 + {\bf k}_2 = 0$.
The relative momentum is ${\bf k} = \frac{1}{2}({\bf k}_1 - {\bf k}_2)$ and it 
 follows that ${\bf k}_1 = {\bf k}$ and ${\bf k}_2 = - {\bf k}$.
The masses of the hadrons are 
$m_1$ and $m_2$, and the energies are $E_1 = \sqrt{{\bf k}^2 + m_1^2}$ and $E_2 = \sqrt{{\bf k}^2 + m_2^2}$.
For fixed energies of the hadrons, relative momentum, ${\bf k}$, is a vector of 
fixed magnitude, $k$. 

Two equivalent ways of writing the lattice coordinates of the operators are,
\beqa
 {\bf x}_1 = {\bf X}+\frac{1}{2}{\bf x},~~~ {\bf x}_2 = {\bf X}-\frac{1}{2}{\bf x},
\eeqa
and 
\beqa
{\bf X} = \frac{1}{2}\Big({\bf x}_1 + {\bf x}_2\Big), ~~~~{\bf x} = {\bf x}_1 - {\bf x}_2.
\eeqa
We switch between them  without further comment.

Consider a product of hadron operator, $A_{s_1,m_{s_1}}({\bf x}_1,t)$, 
and hadron operator, $B_{s_2,m_{s_2}}({\bf x}_2,t)$, where $s_1$ and $s_2$ are intrinsic-spin labels, and 
$m_{s_1}$ and $m_{s_2}$ are labels for the z-components of intrinsic spins. 
The desired parent operator involves coupling intrinsic spins and relative orbital angular momentum 
to total angular momentum, $J$, and projection, $M$, as follows,
\beqa
&& \left( A B \right)_{J^PM}({\bf P}=0,k,t) = 
\!\!\! \int d^3x_1 \!\!\! \int d^3x_2 
\sum_{\substack{s_1,s_2\\{m_{s_1},m_{s_2}}\\{ \ell,m}}} {\cal C}_{JM}
\nonumber \\
&& \times \Big[e^{i {\bf k} \cdot {\bf x}}\Big]_{\ell,m} \!\!\!
   A_{s_1,m_{s_1}}({\bf x}_1,t) B_{s_2,m_{s_2}}({\bf x}_2,t),
\nonumber \\  \label{eq:mB} 
\eeqa
where ${\cal C}_{JM}$ denotes the 
following product of the usual Clebsch-Gordan coefficients for coupling the spins and orbital angular momenta,
\beqa
{\cal C}_{JM} = \langle s_1 m_{s_1} s_2 m_{s_2}| S m_S\rangle \langle S m_S \ell m| J M\rangle,
\eeqa
and the partial wave of relative momentum is given by Eq.~(\ref{eq:PW}).

A matrix of correlation functions for the hadron-hadron system based on parent operators is, 
\beqa
&&   C_{J^{\prime}M^{\prime}:JM}({\bf P}=0,k, t)  = 
 \big{\langle} 0 \big{|} \left( A B \right)_{J^{\prime},M^{\prime}}({\bf P}=0,k,t) \nonumber \\
&& \times \overline{\left(A B \right)}_{JM}({\bf P}=0,k,0) \big{|}0 \big{\rangle}.
\label{eq:ABAB}\eeqa

   Sets of mutually-orthogonal parent operators for two-hadron scattering states are given
in Table~\ref{tab:JMops-S=zero} for total intrinsic spin $S=0$, in Table~\ref{tab:JMops-S=half} 
for $S=\frac{1}{2}$ and Table~\ref{tab:JMops-S=one} for $S=1$. The tables provide 
the quantum numbers corresponding to nonzero Clebsch-Gordan coefficients for the
 coupling of intrinsic spins with orbital angular 
momenta $\ell$ = 0, 1, 2, 3 and 4. Some of the listed operators are orthogonal to others
because their couplings involve pairs of $\ell, m$ values that are orthogonal, or pairs of $m_s$
values that are orthogonal. Other operators, for example $\frac{3}{2}^+\!\!,\frac{3}{2}$ and $\frac{5}{2}^+\!\!,\frac{3}{2}$, 
involve combinations of Clebsch-Gordan coefficients that give orthogonality. Note that 
each $\ell, m$ component involved must be normalized to the same value in order 
to realize the orthogonality due to the Clebsch-Gordan 
coefficients.$^{\footnotemark[1]}$
\footnotetext[1]{It is straightforward to use the orthogonality
properties of C-G coefficients to project out the $\ell, m$ components corresponding to a given $m_s$.}   

It is straightforward to add operators for other momenta in order to build a 
variational basis of operators with respect to energy within each set 
of $J$, $M$ values.  For each additional momentum included, all operators of
 Table~\ref{tab:JMops-S=zero} ( or Table~\ref{tab:JMops-S=half} or 
Table~\ref{tab:JMops-S=one}) can be used. Orthogonality is then
between subsets of operators that have different $J$, $M$ values. 

Isospin and the Pauli Principle provide limits on the partial wave operators that 
should be used.  For example, the $\pi\pi$ system can occur in states of isospin
0, 1 or 2 and the system must be symmetric with regard to the exchange of the pions. 
Similarly, the $NN$ system can occur in states of isospin 0 or 1 and the system must be 
antisymmetric with regard to exchange of the two nucleons. Operators that do not obey the correct symmetry should be omitted.


\begin{table} 
\caption{ Mutually-orthogonal parent operators for the 
scattering of two hadrons with total intrinsic spin $S=0$, for example, $\pi \pi$
or $NN$. 
The final column provides conventional labels of the form $^{2S+1}L_J$ for the 
$NN$ system with $S=0$.\label{tab:JMops-S=zero}}
  \begin{tabular}{|c|c|c|c|}
\hline
Operator& $J^P$,$M$  &  $\ell$,$m$ &  $^{2S+1}L_J$ \\ 
\hline
 1      &  $0^+,0$  &  0,0      &  $^1S_0$  \\
\hline
 2      &  $1^-,1$  &  1,1      &  $^1P_1$  \\
\hline
 3      &  $1^-,0$  &  1,0      &  $^1P_1$  \\
\hline
 4      &  $2^+,2$  &  2,2      &  $^1D_2$  \\
\hline
 5      &  $2^+,1$  &  2,1      &  $^1D_2$  \\
\hline
 6      &  $2^+,0$  &  2,0      &  $^1D_2$  \\
\hline
 7      &  $3^-,3$  &  3,3      &  $^1F_3$  \\
\hline
 8      &  $3^-,2$  &  3,2      &  $^1F_3$  \\
\hline
 9      &  $4^-,3$  &  4,3      &  $^1G_4$  \\
\hline 
\end{tabular}
\end{table}

\begin{table} 
\caption{ Mutually-orthogonal parent operators for the scattering of two hadrons
with total intrinsic spin $S = \frac{1}{2}$, for example, $\pi N$.  
Each row of the table gives the combinations of $\ell,m$ and $m_s$
quantum numbers that occur in the Clebsch-Gordan expansion to form $J^P$ and $M$. 
 \label{tab:JMops-S=half}}
  \begin{tabular}{|c|cc|cc|cc|cc|}
\hline
Operator& $J^P$     &  M            &$\ell,m$& $m_s$       &$\ell,m$&  $m_s$  \\
\hline
1 & $\frac{1}{2}^+$ & $\frac{1}{2}$ &  0,0  &$\frac{1}{2}$ &       &                 \\
\hline
2 & $\frac{1}{2}^-$ & $\frac{1}{2}$ &  1,0  & $\frac{1}{2}$&  1,1  & -$\frac{1}{2}$  \\
\hline
3 & $\frac{1}{2}^-$ &-$\frac{1}{2}$ &  1,-1 & $\frac{1}{2}$&  1,0  &  -$\frac{1}{2}$  \\
\hline
4 & $\frac{3}{2}^-$ & $\frac{3}{2}$ &  1,1  &$\frac{1}{2}$ &       &                 \\
\hline
5 & $\frac{3}{2}^-$ & $\frac{1}{2}$ &  1,0  &$\frac{1}{2}$ &  1,1  &  -$\frac{1}{2}$ \\  
\hline
6 & $\frac{3}{2}^+$ & $\frac{3}{2}$ &  2,1  & $\frac{1}{2}$&  2,2  &  -$\frac{1}{2}$ \\
\hline
7 & $\frac{3}{2}^+$ & $\frac{1}{2}$ &  2,0  & $\frac{1}{2}$&  2,1  &  -$\frac{1}{2}$ \\  
\hline
8 & $\frac{5}{2}^+$ & $\frac{5}{2}$ &  2,2  &$\frac{1}{2}$ &       &                 \\  
\hline
9 & $\frac{5}{2}^+$ & $\frac{3}{2}$ &  2,1  & $\frac{1}{2}$&  2,2  &  -$\frac{1}{2}$ \\
\hline
10& $\frac{5}{2}^+$ & $\frac{1}{2}$ &  2,0  & $\frac{1}{2}$&  2,1  &  -$\frac{1}{2}$ \\
\hline 
11& $\frac{5}{2}^-$ & $\frac{5}{2}$ &  3,2  &$\frac{1}{2}$ &  3,3  &  -$\frac{1}{2}$ \\  
\hline
12& $\frac{7}{2}^-$ & $\frac{5}{2}$ &  3,2  & $\frac{1}{2}$&  3,3  &  -$\frac{1}{2}$ \\
\hline
13& $\frac{7}{2}^+$ & $\frac{7}{2}$ &  4,3  & $\frac{1}{2}$&  4,4  &  -$\frac{1}{2}$ \\
\hline
\end{tabular}
\end{table}

\begin{table} 
\caption{ Mutually-orthogonal parent operators for the scattering of two hadrons
with total intrinsic spin $S = 1$, for example, $NN$.  
Each row of the table gives the combinations of $\ell,m$ and $m_s$
quantum numbers that have nonzero coefficients
in the Clebsch-Gordan expansion to form $J$ and $M$. 
The final column provides conventional labels of the form $^{2S+1}L_J$ for the
$NN$ system with $S=1$.
 \label{tab:JMops-S=one}}
  \begin{tabular}{|c|cc|cc|cc|cc|c|}
\hline
Operator&$J$ & M  &$\ell,m$& $m_s$ & $\ell,m$& $m_s$ &$\ell,m$&  $m_s$ & $^{2S+1}L_J$ \\
\hline 
1 &    0   &   1    &  0,0   &  1    &         &       &        &        &  $^3S_1$    \\
\hline
2 &    0   &   0    &  0,0   &  0    &         &       &        &        &  $^3S_1$    \\
\hline
3 &    0   &  -1    &  0,0   & -1    &         &       &        &        &  $^3S_1$    \\
\hline
4 &    0   &   0    &  1,1   & -1    &  1,0    &   0   &  1,-1  &  1     &  $^3P_0$    \\
\hline
5 &    1   &   1    &  1,1   &  0    &  1,0    &   1   &        &        &  $^3P_1$    \\
\hline
6 &    1   &   0    &  1,1   &  -1   &  1,-1   &  1    &        &        &  $^3P_1$    \\
\hline
7 &    2   &   2    &  1,1   &  1    &         &       &        &        &  $^3P_2$    \\
\hline
8 &    2   &   1    &  1,1   &  0    &  1,0    &   1   &        &        &  $^3P_2$    \\
\hline
9 &    2   &   0    &  1,1   &  -1   &  1,0    &   0   &  1,-1  &  1     &  $^3P_2$    \\
\hline
10 &   1   &   1    &  2,2   &  -1   &  2,1    &   0   &  2,0   &  1     &  $^3D_1$    \\
\hline
11 &   1   &   0    &  2,1   &  -1   &  2,-1   &   1   &        &        &  $^3D_1$    \\
\hline
12 &   2   &   2    &  2,2   &  0    &  2,1    &   1   &        &        &  $^3D_2$    \\
\hline
13 &   2   &   1    &  2,2   &  -1   &  2,1    &   0   &  2,0   &  1     &  $^3D_2$    \\
\hline
14 &   2   &   0    &  2,1   &  -1   &  2,-1   &   1   &        &        &  $^3D_2$    \\
\hline
15 &   3   &   3    &  2,2   &  1    &         &       &        &        &  $^3D_3$    \\
\hline
16 &   3   &   2    &  2,2   &  0    &  2,1    &   1   &        &        &  $^3D_3$    \\
\hline
17 &   3   &   1    &  2,2   &  -1   &  2,0    &   1   &  2,0   &   1    &  $^3D_3$    \\
\hline
\end{tabular}
\end{table}
 
\subsection{Descendant operators in lattice irreps for ${\bf P} = 0$ \label{subsec:subduce_P=0}}

When the total momentum of a two-hadron operator
is zero, the lattice is invariant with respect to the elements of the octahedral group. 
The parent operators given above are subduced to 
irreps of the octahedral group, which produces descendant operators that can be used in lattice 
calculations.

The subduction coefficients, $S^{J,M}_{\Lambda, r}$, are given in Ref.~\cite{Dudek:2010wm}
for integer spins and in Ref.~\cite{Edwards:2011jj} for half-integer spins.
They should be applied as follows to the
operators of Eq.~(\ref{eq:mB}) that are labeled by $J,M$ to obtain operators that are labeled by
the octahedral group irrep $\Lambda$  and row $r$,
\beq
 \left( A B \right)_{\Lambda,r}({\bf k}_1,{\bf k}_2,t) = \sum_M S^{J,M}_{\Lambda, r}\left( A B \right)_{JM}({\bf k}_1,{\bf k}_2,t). 
\eeq

As shown in appendix B of Ref.~\cite{Edwards:2011jj}, the subduction coefficients are real and they obey,
\begin{eqnarray}
  \sum_M {\cal S}^{J, M}_{\Lambda^{\prime}, r^{\prime}}{\cal S}^{J, M}_{\Lambda, r}  = 
  \delta_{\Lambda^{\prime},\Lambda} \delta_{r^{\prime},r}.
  \label{eq:sum_M_SS}
\end{eqnarray}
This property causes descendant states to be orthogonal with respect to $J$ 
when they are obtained by subducing a set of mutually orthogonal parent states.
This is shown for quantum states of angular momentum, $|J,M \rangle$, and their descendant
states, labelled by lattice irreps, that are subduced from a single value of $J$, i.e.,
$| \Lambda[J], r\rangle = \sum_{M}{\cal S}^{J, M}_{\Lambda, r} | J,M\rangle $, as follows,
\beqa
\langle \Lambda^{\prime}[J^{\prime}], r^{\prime} | \Lambda[J], r\rangle &=& 
\sum_{M,M^{\prime}} {\cal S}^{J^{\prime}, M^{\prime}}_{\Lambda^{\prime}, r^{\prime}} 
 {\cal S}^{J, M}_{\Lambda, r} \langle J^{\prime},M^{\prime} | J,M \rangle, \nonumber \\
&=& \sum_{M,M^{\prime}} {\cal S}^{J^{\prime}, M^{\prime}}_{\Lambda^{\prime}, r^{\prime}}             
 {\cal S}^{J, M}_{\Lambda, r} \delta_{J^{\prime},J}\delta_{M^{\prime},M} \nonumber \\
&=& \delta_{ \Lambda^{\prime},\Lambda}\delta_{ r^{\prime},r}\delta_{J^{\prime},J}.
\label{eq:irrep-orthogonality}
\eeqa

Spin identification is based on an orthogonal set of parent 
operators.  
A matrix of correlation functions is calculated based on the descendant, irreducible operators. 
Because of the orthogonality of Eq.~(\ref{eq:irrep-orthogonality}), the matrices are approximately 
block-diagonal with respect to the values of the parent spin, $J$, from 
which the irreducible operators are subduced.   
Use of the variational method\cite{Michael:1985ne,Luscher:1990ck} to extract the energies of states, and operator overlaps~\cite{Dudek:2010wm,Edwards:2011jj}, 
$Z_{nm}= \langle n | O_{m} | 0\rangle$,
to identify the couplings of operators to the states, then identifies which 
parent spins, $J$, are predominant in the creation of a lattice state.
Generally, when a spin-$J$ operator couples to a state, that identifies 
$J$ as a lower bound of the spin. 
If two or more operators couple to a state, the highest value of $J$ identifies the lower bound of the spin. 

The phase shift can be calculated 
using the Luscher method, which is based on the deviation of the energy
of a scattering state from the energy when there is zero interaction.  The 
latter energy may be calculated from the masses of the hadrons and the
quantized momenta that obey periodic boundary conditions.

\subsection{Two-hadron operators with total momentum  ${\bf P}\neq 0$ \label{subsec:P_not_0}}

Nonzero total momentum is of considerable interest because it 
provides a way to increase the number of c.m. energies
that can be obtained using a given lattice.

The helicity, $\lambda$, is the eigenvalue of $J_z$ in the c.m. frame.
Invariance with respect to boosts along the direction of the total momentum
shows that helicity takes the same value in the c.m. frame and the lattice frame,
where it can be determined. 

Because $J \geq |\lambda|$, 
the use of operators for a range of helicity values provides the following signature.
A state with spin $J$ couples to the operators with $\lambda \leq J$, and not 
to the operators for higher helicity.  

It is useful to decompose coordinates and momenta into parts that are perpendicular and parallel to 
${\bf P}$, which are denoted by subscripts, $\perp$ and $\parallel$, respectively. 
Rotations that keep ${\bf P}$ invariant also leave the parallel components of other vectors invariant, e.g.,
$k_{1\parallel}$, $k_{2\parallel}$ and $k_{\parallel} = \frac{1}{2}\Big(k_{1\parallel} -k_{2\parallel}\Big)$. 
Vectors that are perpendicular to the total momentum, e.g.,
${\bf k}_{1\perp} = -{\bf k}_{2\perp} = {\bf k}_{\perp}$, are invariant with respect to
boosts along the direction of the total momentum. 
Energies are $E_1 = \sqrt{m_1^2 + k^2_{\perp} + k^2_{1 \parallel}}$ and 
$E_2 = \sqrt{m_2^2 + k^2_{\perp} + k^2_{2 \parallel}}$, where $k_{\perp} = |{\bf k}_{\perp}|$ is invariant 
with respect to rotations that keep ${\bf P}$ invariant.

\subsection{ Parent helicity operators for ${\bf P} \neq 0$ \label{subsec:P_not_0_parent}}

  Two-hadron operators are labelled by the total helicity
when ${\bf P} \neq 0$.   
For two spin-zero hadrons, the helicity values are constructed in the lattice frame
using a cylindrical function in place of the spherical harmonic 
that appears in Eq.~(\ref{eq:PW}), as follows,
\beq
\Big[ e^{i {\bf k}_{\perp} \cdot {\bf x}_{\perp}}\Big]_m =   J_m(k_{\perp}x_{\perp}) e^{i m \phi_x},
\eeq
where $\phi_x$ is the azimuthal angle of ${\bf x}_{\perp}$ and $J_m$ is a Bessel function.  
An overall factor $i^m e^{-i m \phi_k}$ is omitted, where $\phi_k$ is the azimuthal angle of
${\bf k}_{\perp}$, because it plays no role in the final analysis.

A two-hadron parent operator incorporates intrinsic spins of the hadrons as follows,
\beqa
&&\left( A B \right)_{\lambda}({\bf P},k_{\parallel}, k_{\perp},t) = 
\!\! \int \!\! d^3x_1 \!\! \int \!\! d^3x_2 e^{i {\bf P} \cdot {\bf X}} e^{i  k_{\parallel} x_{\parallel}}
\!\!\!\!\!\!\! \sum_{\substack{s_1,s_2\\{m_{s_1},m_{s_2},m}}} 
\nonumber \\
&&{\cal C}_{\lambda} ~ \Big[ e^{i {\bf k}_{\perp} \cdot {\bf x}_{\perp}}\Big]_m ~ 
   A_{s_1,m_{s_1}}({\bf x}_1,t) B_{s_2,m_{s_2}}({\bf x}_2,t),
  \label{eq:mB_Cm} 
\eeqa
where ${\cal C}_{\lambda}$ denotes the 
following product of a Clebsch-Gordan coefficient for spin and a delta function,
\beqa
{\cal C}_{\lambda} = \langle s_1 m_{s_1} s_2 m_{s_2}| S m_S\rangle ~ \delta_{m_S+m,~\lambda}.
\eeqa
This operator is constructed to create a state with helicity, $\lambda$, that is 
the sum of the orbital helicity, $m$, and spin helicity, $m_S$.

The matrix of correlation functions for the two-hadron system based on parent operators is, 
\beqa
&&   C_{\lambda^{\prime};\lambda}({\bf P},k_{\parallel},k_{\perp},t)  = 
 \big{\langle} 0 \big{|} \left( A B \right)_{\lambda^{\prime}}({\bf P},k_{\parallel},k_{\perp},t) \nonumber \\
&& \times \overline{\left(A B \right)}_{\lambda}({\bf P},k_{\parallel},k_{\perp},0) \big{|} 0 \big{\rangle}.
\label{eq:ABAB_Cm}
\eeqa

\subsection{Descendant operators in lattice irreps for ${\bf P} \neq 0$ \label{subsec:P_not_0_irreps}}
 
When the total momentum of the two-hadron system is nonzero, the lattice symmetry group is  
$C_{nv}$ (alternatively known as $Dic_n$ for the double group), for $n$ =4, 3, or 2.  For $C_{4v}$,
the total momentum is directed along one of the axes of the cube; for $C_{3v}$, 
it is along a body diagonal; for $C_{2v}$ it is parallel to a face diagonal. 
For each of these cases, we label the direction of total momentum as the z-axis.
Altmann and Herzig \cite{Altmann:1994} have developed the subductions from
$SU(2)$ to irreps of the point groups $C_{nv}$; their notation is followed in this work.
Equivalent results
are given in Refs.~\cite{Thomas:2011rh,Moore:2006a,Moore:2006b}. 

The $C_{nv}$ group applies for integer spins and the $C_{nv}$ double group applies 
for half-integer spins. The group elements are $n$ rotations about the z direction, $R_z(\theta)$,
with equally spaced $\theta$ values in the range $0 \leq \theta < 2\pi$ (or $2n$ rotations in the range 
$0 \leq \theta < 4\pi$ for the double group).  In addition, there is
an equal number of reflections, $R_z(\theta) \Pi_{yz}$, that involve the same rotations and a 
reference reflection, which is chosen to be $\Pi_{yz}$, the reflection in the
$yz$-plane. The group elements transform helicity states as follows,
\beqa
R_z(\theta)\bigr|\lambda \bigr\rangle = e^{-i \lambda \theta} \bigr|\lambda \bigr\rangle , 
\nonumber \\
\Pi_{yz} \bigr|\lambda \bigr\rangle  =  \eta \bigr|-\lambda \bigr\rangle,
\label{eq:Rz_Pi}
\eeqa
 where $\eta  = (-1)^{J} P$ and $P$ is the parity.
The phase factor, $\eta$, is equal to $\tilde{\eta}$ in the
analysis of Ref.~\cite{Thomas:2011rh}, where it is shown to follow from the choice of 
$\Pi_{yz}$ as the reference reflection,
 
Representations of $C_{nv}$ involve a linear combination of helicity states,
$\bigr|\lambda \bigr\rangle$ and $\bigr|-\lambda\bigr\rangle$.
In this work, they are chosen to diagonalize the reference
reflection  and are 
 labeled by the positive helicity, $\lambda > 0 $, 
and a sign, $\pm$,  which serves as the row index of the representation, as follows,
\beqa
\bigr| \lambda, \pm \bigr\rangle \equiv \sqrt{\frac{1}{2}} \Big(~\bigr| \lambda \bigr\rangle ~\pm  \eta \bigr| -\lambda ~\bigr\rangle \Big).
\label{eq:Cnv_rows}
\eeqa
The action of group elements is to transform the $ \bigr| \lambda, \pm \bigr\rangle$ 
representations into linear combinations of one another, 
\beqa
 R_z(\theta) \bigr| \lambda, + \bigr\rangle &=& cos(\lambda \theta) \bigr| \lambda, +\bigr\rangle -i sin(\lambda \theta) \bigr| \lambda, -\bigr\rangle, 
\nonumber \\
 R_z(\theta) \bigr| \lambda, -\bigr\rangle &=& -i sin(\lambda \theta) \bigr| \lambda, +\bigr\rangle + cos(\lambda \theta) \bigr| \lambda, -\bigr\rangle,
\nonumber \\
 \Pi_{yz} \bigr| \lambda, \pm\bigr\rangle &=& \pm (-1)^J \bigr| \lambda, \pm\bigr\rangle,
\label{eq:Cnv_action}
\eeqa 
where the eigenvalue of the reference reflection is seen to be $ \pm (-1)^J$.  
Note that a two-dimensional representation is reducible to two
one-dimensional irreps when 
 $sin(\lambda \theta)$ vanishes for each group rotation angle, $\theta$.
Otherwise $\bigr| \lambda,+ \bigr\rangle$ and 
$\bigr| \lambda,- \bigr\rangle$ provide the rows of a two-dimensional irrep.

Descendant operators for ${\bf P} \neq 0$ transform as irreps of $C_{nv}$.
They involve the positive helicity, $\lambda > 0$, and the row index, $\pm$, 
and have a composition similar to Eq.~(\ref{eq:Cnv_rows}),
\beqa
(AB)_{\lambda, \pm}({\bf P},k_{\parallel},k_{\perp},t) = \sqrt{\frac{1}{2}} 
\Big( (AB)_{\lambda}({\bf P},k_{\parallel},k_{\perp},t) 
\nonumber \\
\pm  \eta (AB)_{-\lambda}({\bf P},k_{\parallel},k_{\perp},t)\Big).
\label{eq:mB_Cnv}
\eeqa

The rows of a two-dimensional irrep are orthogonal and they produce the same 
correlation function as the average over rows, which is the same as the average over $\pm$ signs.  
Using the descendant operators 
of Eq.~(\ref{eq:mB_Cnv}), the average over rows is,
\beq
\overline{C}_{\lambda} = \frac{1}{2} \left( C_{\lambda,+;\lambda,+} + C_{\lambda,-;\lambda,-} \right) .
\eeq

A little algebra shows that this is equal to the average over 
the helicities, $\lambda$ and $-\lambda$, that appear when the expression above
is expanded in terms of the correlation functions $C_{\lambda^{\prime},\lambda}$ of Eq.~(\ref{eq:ABAB_Cm}).
Using $\eta^*\eta = 1$, the row average is evaluated as follows,
\beqa
\overline{C}_{\lambda} 
    &&       = \frac{1}{4} \left( C_{\lambda,\lambda} + \eta^* C_{-\lambda,\lambda} 
+ \eta C_{\lambda,-\lambda} +
             \eta^* \eta   C_{-\lambda,-\lambda} \right) 
\nonumber \\
    &&       + \frac{1}{4} \left( C_{\lambda,\lambda,} - \eta^* C_{-\lambda,\lambda} - \eta C_{\lambda,-\lambda} +
               \eta^* \eta  C_{-\lambda,-\lambda} \right) ,
\nonumber \\
    &&       = \frac{1}{2} \left(C_{\lambda,\lambda} + C_{-\lambda,-\lambda}\right).
\label{eq:C_avg}
\eeqa
The parity-dependent phase factor, $\eta$, cancels out of the average over rows.

Each one-dimensional irrep also involves both $\lambda$
and $-\lambda$, as in Eq.~(\ref{eq:Cnv_rows}). For most purposes,
it is adequate to average the two one-dimensional 
irreps for a given helicity, $\lambda$, in the same manner as the rows of a two-dimensional irrep.  
When this is done, the result is the same as
in Eq.~(\ref{eq:C_avg}), i.e., the parity-dependent phase factor, $\eta$, 
cancels out.  Thus, correlation functions can be calculated 
directly from parent operators, as in Eq.~(\ref{eq:ABAB_Cm}), 
and then averaged over the $\pm \lambda $ values in order to obtain the same
result as would be obtained from averaging over the rows of correlation functions based on 
the irrep operators of Eq.~(\ref{eq:mB_Cnv}). 

Although the average over helicities provides a simple and straightforward 
analysis, the irrep operators are given in Appendix~\ref{sec:Cnv_irreps} for completeness.



The orthogonality of periodic orbital-helicity functions involves the following
matrix of overlaps,
\beqa
&&Z_{m^{\prime} m} = \int d^3 x 
\big[ J_m(k_{\perp}x_{\perp}) e^{i m \phi_x}\big]^{(per) \dag} \times \nonumber \\
&&\!\!\!\!\!\!  \big[J_{m^{\prime}}(k_{\perp}x_{\perp})e^{ i m^{\prime} \phi_x }\big]^{(per)} .
\label{eq:m'm-orthog}
\eeqa
This matrix is normalized as in Eq.~(\ref{eq:Z-lm-orthog}) in order to make diagonal elements 
equal to one. Then the zero elements depend on index $n$ of group $C_{nv}$ as follows,
\beq
 Z_{m^{\prime},m} = 0 ~~if~m^{\prime}~\neq~m~(mod~n).
\eeq
Table~\ref{tab:Zmm} shows the pattern of matrix elements based on Eq.~(\ref{eq:m'm-orthog}).

\begin{table} 
\caption{ Pattern of matrix elements $Z_{m^{\prime} m}$ for group $C_{4v}$. 
For group $C_{3v}$, the x's shift to horizontal positions that are a multiple of
three units away from the diagonal and for group $C_{2v}$ the x's move to positions 
that are a multiple of 2 
units away from the diagonal.  Symbols
have the same meaning as in Table~\ref{tab:even-L-orthog}. The top row shows the $ m$ indices and the 
left column shows the $m^{\prime}$ indices. 
\label{tab:Zmm} }
  \begin{tabular}{l|c|c|c|c|c|c|c|c|c|c|}
     &~-4~&~-3~&~-2~&~-1~&~\ 0~&~\ 1~&~\ 2~&~\ 3~&~\ 4~&\ \\
   \hline
\hline
  -4 & \1&   &   &   & x  &   &   &   & x   \\
\hline
  -3 &   & \1&   &   &    & x &   &   &     \\
\hline
  -2 &   &   & \1&   &    &   & x &   &     \\
\hline
  -1 &   &   &   &\1 &    &   &   & x &     \\
\hline
   0 & x &   &   &   & \1 &   &   &   & x   \\
\hline
   1 &   & x &   &   &    & \1&   &   &     \\
\hline
   2 &   &   & x &   &    &   &\1 &   &     \\
\hline
   3 &   &   &   & x &    &   &   &\1 &     \\
\hline
   4 & x &   &   &   &  x &   &   &   &\1   \\
\hline
\hline
\end{tabular}
\end{table}

The best case for spin identification is group $C_{4v}$ because it has the largest range of 
orbital helicities that can be included in an orthogonal set. 
The orbital-helicity
states that are orthogonal are limited to $m$ = 0 and $\pm 1$ for both $C_{3v}$ and $C_{2v}$. 
For each of these groups, the resulting sets of 
mutually-orthogonal operators for two-hadron scattering states are given
in Table~\ref{tab:lambda_orthog-S=zero} for total intrinsic spin $S=0$, 
in Table~\ref{tab:lambda_orthog-S=half} for $S=\frac{1}{2}$ and 
Table~\ref{tab:lambda_orthog-S=one} for $S=1$. 
As each irrep operator includes helicity values, $\lambda$ and $-\lambda$, quantum 
numbers $m$ and $-m$ are included for the cylindrical harmonics. 
Orthogonality for $S=0$
is limited because each $m$-component must be orthogonal to all the other $m$-components in the table.
For example, $m = \pm 3$ cannot be included because $m=\pm 1$ is included.

When intrinsic spin states are included, orthogonality
depends on each $m,m_s$ pair being orthogonal to each other pair. The orthogonality of spins
 allows more operators to be included. 

\begin{table} 
\caption{ Mutually orthogonal helicity operators for group $C_{4v}$ and total intrinsic spin $S = 0$.
Each row lists the irrep label, $ \lambda $, and the $m$ values included
in the operator. For groups $C_{3v}$ and $C_{2v}$, only rows 1 and 2 involving $m$ = 0 and $\pm 1$ 
are orthogonal.  \label{tab:lambda_orthog-S=zero}}
  \begin{tabular}{|c|c|c|c|}
\hline
  No. & ~$\lambda$~ &~m~&~-m~\\
   \hline
   1  &   0   & 0    &  0     \\
\hline
   2  &   1   & 1    & -1     \\
\hline
   3  &   2   & 2    & -2     \\
\hline
\end{tabular}
\end{table}

\begin{table} 
\caption{ Mutually orthogonal helicity operators for group $C_{4v}$ and total intrinsic spin $S = \frac{1}{2}$.
Each row lists the irrep label, $\lambda$, and the $m,m_s$ values
included in the operator.   For groups $C_{3v}$ and $C_{2v}$, only the rows involving $m$ = 0 and $\pm 1$
are orthogonal. \label{tab:lambda_orthog-S=half}}
  \begin{tabular}{|c|c|c|c|}
\hline
  No. & ~$\lambda$~ &~m,$m_s$~&~-m,-$m_s$~\\
   \hline
   1  & \half     & 0 ,    \half & 0  ,     -\half \\
\hline
   2  & \half     & 1 ,   -\half & -1  ,     \half \\
\hline
   3  & \threehalf& 1 ,    \half & -1 ,     -\half \\
\hline
   4  & \fivehalf& 2 ,    \half & -2 ,     -\half \\
\hline
\end{tabular}
\end{table}

\begin{table} 
\caption{ Mutually orthogonal helicity operators for group $C_{4v}$ and total intrinsic $S = 1$.
Each row lists the irrep label, $\lambda$, and the $m,m_s$ values included 
in the operator. For groups $C_{3v}$ and $C_{2v}$, only the rows involving $m$ = 0 and $\pm 1$
are orthogonal.   \label{tab:lambda_orthog-S=one}}
  \begin{tabular}{|c|c|c|c|}
\hline
  No. & ~$\lambda$~ &~$m,m_s$~&~-$m,-m_s$~\\
   \hline
   1  &   0   & 0,0    &  0,0     \\
\hline
   2  &   0   & 1,-1    & -1,1     \\
\hline
   3  &   1   & 0,1    &  0,-1     \\
\hline
   4  &   1   & 1,0    & -1,0     \\
\hline
   5  &   1   & 2,-1   &  -2,1    \\
\hline
   6  &   2   & 1,1    &  -1,-1   \\
\hline
   7  &   2   & 2,0    &  -2,0    \\
\hline
\end{tabular}
\end{table}

\section{Summary \label{sec:summary}}
Operators are developed to facilitate the identification of spins in lattice QCD calculations of
the scattering of two hadrons. In the case that the total momentum of the hadrons is ${\bf P} = 0$, 
partial-wave operators are used to create periodic partial waves of orbital angular momentum, $\ell$,
and projection, $m$.  The operators are based on momenta of magnitude
$k = \frac{2 \pi |{\bf n}}{L}$, where is ${\bf n}$ is a vector with integer components.
These are the momenta for noninteracting plane waves on the cubic lattice, however, other values of $k$ could
be used to provide a different radial smearing of the operators.
Phase shifts may be obtained using the L\"{u}scher method and its extensions.
The periodic partial waves are orthogonal with respect to $\ell$ and $m$, subject to 
restrictions imposed by the cubical lattice. The partial waves provide an advantage relative to the 
use of combinations of plane waves in that there is freedom to select sets of orthogonal
operators based on specific $\ell$ and $m$ values.

The essential point of this work is to differentiate between lattice irreps
that are subduced from different continuum spins. 
For example, the use of mutually orthogonal $\ell$ = 1 and $\ell$ = 3 parent operators 
based on partial waves provides orthogonal descendant operators after subduction to the $T_1$ lattice irrep.
Strong coupling of a lattice state to one of these operators but not the
other provides a signal for the spin of the state. 
While contributions of $\ell \ge$ 5 can occur in the subduced operators,
the short range of hadronic interactions and the centrifugal barrier at high $\ell$
suggest that the role of higher spins is suppressed.

Parent operators for total spin, $J\leq 4$, and projection, $M$, are obtained by coupling
the  intrinsic spins of the hadrons and the partial waves of orbital angular momentum.
Sets of mutually-orthogonal parent operators corresponding to different values of $J$ and $M$
are given for total intrinsic spin $S$ = 0, $\frac{1}{2}$ and 1 based on omitting 
operators that are not orthogonal because of the cubic symmetry.  
 When these parent operators are subduced to lattice irreps, their descendants 
retain orthogonality and lead to matrices of 
correlation functions that are block diagonal with respect to the different values of $J$.  
That allows the spin of a scattering state to be identified as 
the spin of the parent operators that dominate the creation of the state. 

In the case that the total momentum is ${\bf P} \neq 0$, parent operators are designed to
create states of total helicity by combining spin helicities and orbital helicity based on 
cylindrical harmonics.
Mutually-orthogonal sets of parent operators for a sufficient range of total helicity 
values can be used to facilitate the identification of the spin, $J$, of a state.  
The signature is that a state of spin $J$ couples to operators for all total 
helicity values $\lambda \leq J$, and not to operators for $\lambda > J$.
However, there is a limitation owing to cubic symmetry.
Total momenta that are parallel to an axis of the cubical lattice (symmetry 
group $C_{4v}$) allow the widest range of helicity values and, thus,
provide the best case for spin identification based on helicity. 

For either the partial-wave operators or the helicity operators, 
a variational basis can be constructed by using  
additional copies of the sets of mutually-orthogonal operators 
based on different momentum values. 

\acknowledgements
I thank Christopher Thomas for helpful comments 
based on a careful reading of the manuscript.
This material is based upon work supported by the U.S. Department of Energy Office of 
Science, Office of Nuclear Physics, under Award Number DE-FG02-93ER-40762.”

\appendix

\section{Lattice orthogonality of periodic partial waves.\label{app:l_m_orthog}}
 
A generalization of Eqs.~(\ref{eq:O-lm-orthog}) and (\ref{eq:Z-lm-orthog}) is considered in which
the lattice state is assumed to be a scattering state described by the
L\"{u}scher formalism for various values of the phase shift,
\beqa
&&\!\!\!\!\!\! Z_{\ell^{\prime},m^{\prime};\ell, m}= \frac{1}{N_{\ell^{\prime},
m^{\prime}} N_{\ell, m}}
\nonumber \\
&&\times \int d^3x ~\bigr[j_{\ell^{\prime}}(k|{\bf x}|) 
       Y^{\dag}_{\ell^{\prime},m^{\prime}}({\hat x}\bigr]^{(per)} 
 \big[ \Psi_{\ell}(|{\bf x}|)Y_{\ell,m}({\hat x})\big]^{(per)}, \nonumber \\
\label{eq:lm-orthog-x}
\eeqa
where $\Psi_{\ell}(|{\bf x}|)$ is the radial wave function of the scattering state.  
When the phase shift, $\delta_{\ell}$ is zero, $\Psi_{\ell}(|{\bf x}|)$ reduces to a 
spherical Bessel function, $j_{\ell}({k|{\bf x}|})$. 
 
According to the L\"{u}scher formalism, a scattering state for angular momentum $\ell$ and 
phase shift $\delta_{\ell}$
involves a momentum value, $k_{ext}$, in the exterior region, $r > R$, where $R$ is the range of interaction.
Momentum $k_{ext}$ is related to the
energy of the scattering state through $W = \sqrt{m_1^2 + k_{ext}^2} + \sqrt{m_2^2 + k_{ext}^2}$,
as shown by L\"{u}scher~\cite{Luscher:1990ux}.
The value of $k_{ext}$ also is related to the phase shift as follows,
\beqa
\delta_{\ell} = j \pi - \phi(q)  ,\hspace{0.3in}     q = \frac{k_{ext} L}{2\pi},
\label{eq:k_ext}
\eeqa
where the function $\phi(q)$ is obtained from 
\beqa
cot(\phi)= - \frac{{\cal Z}_{00}(1;q^2)}{ \pi^{3/2} q} = -\frac{1}{2 \pi^2 q}\sum_{{\bf n} \in Z^3} 
\frac{1}{({\bf n}^2 - q^2)}.  
\label{eq:phi}
\eeqa
The integer $j$ is chosen so that $\phi(q)$ is a continuous function of $q$ and obeys the boundary condition 
$\phi(0)=0$. 
These formulas neglect contributions from $\ell$ = 4 or higher partial waves.

Since the interaction is not known, it is necessary to use a model to approximate the
scattering wave function in the interior region, $0<r<R$.  
A square-well potential gives a momentum, $k_{int}$, in the
interior region and the scattering wave function is,
\beqa
\Psi_{\ell}(r) \!\!&=&\!\!  j_{\ell}(k_{int} r) , \hspace{1.55in}  0<r<R, \nonumber \\ 
\Psi_{\ell}(r) \!\!&=&\!\!  A \big[cos\delta_{\ell}j_{\ell}(k_{ext} r) - sin\delta_{\ell} n_{\ell}(k_{ext} r)\big],  R<r<L/2,
\nonumber \\ \label{eq:psi}
\eeqa 
where $j_{\ell}$ and $n_{\ell}$ are spherical Bessel functions.

The scattering wave function is then obtained by choosing a phase shift, finding $k_{ext}$ from
Eq.~(\ref{eq:k_ext}) and matching the interior and exterior wave functions in Eq.~(\ref{eq:psi}), and their derivatives, 
at $r=R$ to obtain momentum $k_{int}$ and coefficient A. 

Calculations have used $R=L/4$ as the range of the interaction and $\delta_{\ell}/\pi = [ 0.0, 0.1, 0.25, 0.5, 0.75, 0.9]$.
 The lowest values of $k_{ext}$ are used.  
The cubical volumes considered are $16^3$ and $32^3.$ The construction of periodic
partial waves was done using N =  5, 7, 9 and 11 for the sum in Eq.(\ref{eq:box_sum}). 
For all cases, the results confirm that matrix elements shown by blank entries in 
Tables~\ref{tab:even-L-orthog} and \ref{tab:odd-L-orthog} have numerical values less than $10^{-15}$ in magnitude
for each considered value of the phase shift.

\section{Irreducible representations for helicity states \label{sec:Cnv_irreps}}

As noted in the text, the two-dimensional representation of Eq.~(\ref{eq:Cnv_rows}) 
is reducible when $sin(\lambda \theta)$ vanishes for all group rotations.
Then, the $\big| \lambda,\pm \big\rangle$ states are not mixed by the group 
rotations, as may be seen from Eq.~(\ref{eq:Cnv_action}).

The reduction occurs when $\lambda$ = 0, $n$/2, $n$, $\cdots$, where $n$ is the index of
group $C_{nv}$.  
Following the Mulliken convention\cite{Mulliken:1955}
for labelling irreps, 
the helicity states $\bigr| \lambda,\pm \bigr\rangle$ provide irreps
as listed in Table~\ref{tab:Cnv_IRs}.   
For $\lambda = 0$ or $n$, the $\big| \lambda, +\big\rangle$ state transforms as the $A_1$ irrep and the 
$\big| \lambda, -\big\rangle$ state transforms as
the $A_2$ irrep.  For $\lambda$ = $n$/2 or 3$n$/2, $\big| \lambda,+\big\rangle$ 
transforms as the $B_1$ irrep and $\big| \lambda,-\big\rangle$ transforms as the $B_2$ irrep.

For $\lambda$ = 1, the $\big| \lambda,\pm\big\rangle$ are mixed by the action of 
group rotations for groups $C_{4v}$ and $C_{3v}$, but not for $C_{2v}$.  For $\lambda$ = 3,
the rotations mix the $\big| \lambda,\pm\big\rangle$ for group $C_{4v}$,
providing two rows of the $E_2$ two-dimensional irrep.  For 
groups $C_{3v}$ and $C_{2v}$, there is no mixig and the $\big| \lambda,\pm\big\rangle$
provide one-dimensional irreps $A_1$ and $A_2$ for $C_{3v}$ and $B_1$ and $B_2$ for $C_{2v}$.  

For half-integer spins and helicities, the irreps are two-dimensional and are given labels 
$E_{\frac{1}{2}}$, $E_{\frac{3}{2}}$ and $E$, as shown in Table~\ref{tab:Cnv_IRs}.  
An equivalent table may be found in Ref.~\cite{Moore:2006b}.

\begin{table}[h] 
\caption{Helicity irreps for $C_{nv}$ groups. 
$\Pi = \pm (-1)^J$ is the eigenvalue of the reference reflection
and $\Pi/(-1)^J$ is the row index, namely, $\pm$.
\label{tab:Cnv_IRs} }
  \begin{tabular}{l|c|c|c|c|}
$\lambda~$&~$\Pi/(-1)^J$~&~$~C_{4v}~$~&~$C_{3v}$~&~$C_{2v}$~ \\
   \hline
\hline
  0        & + & $A_1$  & $A_1$  & $A_1$   \\
\hline
  0        & - & $A_2$  & $A_2$  & $A_2$   \\
\hline
 $\frac{1}{2}$ & $\pm$ &  $E_{\frac{1}{2}}$ & $E_{\frac{1}{2}}$ & $E$ \\
\hline
  1        & + & $E_2$  & $E_2$  & $B_1$   \\
\hline
  1        & - & $E_2$  & $E_2$  & $B_2$   \\
\hline
 $\frac{3}{2}$ & $\pm$ &  $E_{\frac{3}{2}}$ & $B_1 \oplus B_2$ & $E$ \\
\hline
  2        & + & $B_1$  & $E_2$  & $A_1$   \\
\hline
  2        & - & $B_2$  & $E_2$  & $A_2$   \\
\hline
 $\frac{5}{2}$ & $\pm$ &  $E_{\frac{1}{2}}$ & $A_1 \oplus B_1$ & $E$ \\
\hline
  3        & + & $E_2$  & $A_1$  & $B_1$   \\
\hline
  3        & - & $E_2$  & $A_2$  & $B_2$   \\
\hline
 $\frac{7}{2}$ & $\pm$ &  $E_{\frac{1}{2}}$ & $E_{\frac{1}{2}}$ & $E$ \\
\hline
  4        & + & $A_1$  & $E_2$  & $A_1$   \\
\hline
  4        & - & $A_2$  & $E_2$  & $A_2$   \\
\hline
\hline
\end{tabular}
\end{table}

\bibliography{LatticePWAbibliography}
\end{document}